\RequirePackage{fix-cm}
\documentclass[smallextended]{svjour3}       
\smartqed  
\usepackage{graphicx}
\usepackage{caption}
\usepackage{subfigure}
\usepackage{epsfig}
\usepackage{epstopdf}
\usepackage{amsmath}
\usepackage[cmintegrals]{newtxmath}
\usepackage{algorithmic}
\usepackage{array}
\usepackage{multirow}
\usepackage[caption=false,font=normalsize,labelfont=sf,textfont=sf]{subfig}
\usepackage{stfloats}
\usepackage{url}
\usepackage[noadjust]{cite}
\begin{document}

\title{VOP Detection for Read and Conversation Speech using CWT Coefficients and Phone Boundaries}


\author{Kumud Tripathi         \and
        K. Sreenivasa Rao 
}


\institute{Kumud Tripathi, K. Sreenivasa Rao \at
              Department of Computer Science and Engineering, \\
              Indian Institute of Technology Kharagpur,\\
              Kharagpur-721302, West Bengal, India\\
              \email{kumudtripathi.cs@gmail.com}  
           \and
           K. Sreenivasa Rao \at
             \email{ksrao@iitkgp.ac.in}
}

\date{Received: date / Accepted: date}

\maketitle

\begin{abstract}

 In this paper, we propose a novel approach for accurate detection of the vowel onset points (VOPs). VOP is the instant at which the vowel begins in the speech signal. Precise identification of VOPs is important for various speech applications such as speech segmentation and speech rate modification. 
 The existing methods detect the majority of VOPs within 40 ms deviation, and it may not be appropriate for the above speech applications.
 To address this issue, we proposed a two-stage approach for accurate detection of VOPs. 
 At the first stage, VOPs are detected using continuous wavelet transform coefficients, and the position of the detected VOPs are corrected using the phone boundaries in the second stage. The phone boundaries are detected by the spectral transition measure method.
 Experiments are done using TIMIT and Bengali speech corpora. 
 Performance of the proposed approach is compared with two standard signal processing based methods. 
 The evaluation results show that the proposed method performs better than the existing methods.
 
\end{abstract}

\section{Introduction}
 {B}{r}oadly, speech can be divided into two modes, namely, read and conversation modes \cite{Batliner1995, Blaauw1991, Dellwo2015}. In read mode, an individual utters in restricted conditions, for instance, reading of books or TV news. However, in conversation mode, two or more individuals are communicating in an unrestricted condition. In general, conversation speech is spontaneous, informal, unstructured, and unorganized. On the other hand, read speech is relatively less expressive and more organized. Due to high dissimilarity, it's important to study the impact of both the modes on the vowel onset point (VOP) detection.

Vowel onset point represents the start of a vowel in a speech signal. VOP is utilized for various speech applications such as (i) spotting consonant-vowel (CV) units in a speech signal \cite{Prasanna2001,Sarma2017}, (ii) identifying other speech events such as formant transition, burst, aspiration, which help significantly in speech recognition \cite{Hermes1990} (iii) dividing speech into vowel and non-vowel like regions \cite{Deb2017}, etc. 
In general, the speech signals are processed at sub-word level such as monophone and triphone for speech recognition. Gangashetty et al. \cite{Gangashetty2004}, have shown that syllables are the relevant sub-word units for speech recognition in Indian languages. Syllable represents a group of consonants (C) and vowel (V) in the form of $C^aVC^x$, where $a$ and $x$ indicate the count of consonants before and after the vowel, respectively.  Among all the $C^aVC^x$ combinations, the CV units are the most common (about 90\%) existing syllables in Indian languages \cite{Gangashetty2004}.
The CV units can be identified by accurate detection of VOPs in a continuous speech \cite{Gangashetty2005}. Therefore, the performance of the speech recognition system will be affected by the accuracy of the VOP detection method \cite{Prasanna2001}. 
In literature \cite{Hermes1990,Gangashetty2004, Gangashetty2004a,Prasanna2005,Rao2009,Prasanna2009, Vuppala2012}, traditional vowel onset point detection methods are developed for read mode of speech. However, in a realistic scenario, conversation speech is more frequently observed than the read speech \cite{Furui2003,Nakamura2008}. In terms of acoustic and linguistic characteristics, conversation mode has significant variations than the read mode \cite{Furui2003,Nakamura2008}. 
Thus, the existing VOP detection methods may lead to spurious detection as well as missing VOPs for conversation speech. Hence, the performance of the traditional speech recognition system will be drastically reduced for conversation speech. Therefore, it is required to accurately detect the VOPs in conversation and read modes of speech for achieving the better recognition accuracy. 
So, the current study is motivated by the recognition of speech for Indian languages in read and conversation modes.

 In previous studies, various methods are explored for VOP detection based on statistical modeling and signal processing approaches. Different statistical modeling methods are explained in \cite{Gangashetty2004, Wang1991, Gangashetty2004a} which utilizes multilayer feed-forward neural network, hierarchical neural network, and auto-associative neural network for VOP detection. These networks are developed using the speech features extracted from both sides of the VOPs. The predicted frame types are used for detecting VOPs in a speech signal. 
On the other hand, the detection of VOPs using signal processing methods is implemented by deriving various speech features.  
In \cite{Hermes1990}, VOPs are detected by finding the locations of rapid growth in the vowel intensity. The change in the energy of each peak and valley of a speech signal is representing the vowel intensity. 
Prasanna et al., \cite{Prasanna2009} proposed a VOP detection method based on the fusion of evidence from the excitation source, spectral peak energy, and modulation spectrum. The performance of the combined approach is better than the individual methods for VOP detection. Vuppala et al. \cite{Vuppala2012} utilized spectral energy present in glottal closure regions of speech signal for VOP detection. The computed spectral energy is robust and high around the glottal closure instants (GCIs). 

Mostly, the statistical modelling methods and signal processing techniques may falsely detect the VOPs in the presence of diphthongs and semivowel-vowel transitions \cite{Prasanna2011,Kumar2016,Kumar2017}. This is due to the similar acoustic characteristics of the semivowels and vowels.
Hence, in the recent works, various statistical modelling \cite{Sarma2013,Khonglah2014}, and signal processing \cite{Prasanna2011,Deb2017} methods have been evolved for detecting the vowel-like region onset points (VLROP) instead of vowel onset point. The VLROPs represents the start of the vowel, semivowel, and diphthong speech regions \cite{Prasanna2011}. But these methods may not be suitable for speech applications where only vowel regions are needed to be identified, such as consonant-vowel recognition, speech-rate modification, speaker recognition, and so on.

The existing methods based on statistical modelling techniques depend on a huge amount of training data. 
In these methods, at the first step, a classifier is trained for detecting the vowel regions, and then the VOPs are detected by locating the instant at which detected vowels are started. The accuracy of the detected VOPs will depend on the performance of the vowel detection algorithm.
However, the signal processing methods can be directly applied to speech signals for identifying VOPs as compared to statistical modelling methods. The signal processing methods follow simple and less number of steps, then the statistical modelling methods which follow the complex process.
In terms of accuracy, both methods are providing almost similar results. Hence, in this work, we have proposed a signal processing based method for accurate detection of VOPs. As the proposed method is signal oriented, so; the state of the art signal processing methods \cite{Prasanna2009, Vuppala2012} are included for performance comparison.
The existing methods \cite{Prasanna2009, Vuppala2012,Kumar2016, Kumar2017, Prasanna2011} detect the majority of the VOPs within 40 ms deviation. Therefore, attaining a better accuracy for VOP detection at lower deviation is the primary goal of the proposed approach.

In this work, we have proposed a novel method to accurately detect the VOPs in a speech signal. 
The proposed method is performed at two stages for robust detection of VOPs. At the first stage, continuous wavelet transform (CWT) is explored for determining the VOP evidence. Continuous wavelet transform \cite{Stephane1999} is capable of detecting the instants of sharp transitions, including steady regions in a speech signal. This is the motivation behind choosing CWT for VOP detection.  
At the second stage, a new approach is explored based on phone boundary information for correcting the positions of detected VOPs. Spectral transition measure (STM) method \cite{Madhavi2015, Dusan2006, Furui1986} is applied for detecting the phone boundaries. 
Dusan et al. \cite{Dusan2006} have shown that the STM is accurately detecting 90\% of phone boundaries under 20 ms deviation. In this work, it is analyzed that the majority of VOPs detected using CWT coefficients are within 40 ms deviation. Therefore, to improve the accuracy of the proposed method at low deviation, the location of the detected VOPs are corrected with the help of detected phone boundaries. 
The proposed method is significant for segregating the vowel onset points from the remaining speech regions. To validate this fact, the proposed method is compared with signal processing techniques reported in \cite{Prasanna2009, Vuppala2012} using TIMIT corpus. In addition to that, the importance of the proposed method is shown by detecting VOPs for read and conversation modes of Bengali speech. 


The organization of the paper is as follows. Section \ref{base} 
describes the baseline VOP detection methods. The description of the proposed method for accurate VOP detection is presented in Section \ref{prop}. 
The performance and significance evaluations of the proposed method using TIMIT and Bengali (read and conversation modes) speech corpora are presented in Section \ref{eval}. Section \ref{conc}, includes the conclusions of the current study and works that need to be explored in the future.


\section{Baseline VOP Detection Methods}\label{base}
Performance of the proposed approach is compared with two
standard signal processing based methods. The first method
combines the evidence from the excitation source, spectral peaks
energy, and modulation spectrum \cite{Prasanna2009}, and the second method is based on spectral energy around glottal closure regions \cite{Vuppala2012}. The detailed description of these methods are presented below.

\subsection{Combined evidences from excitation source, spectral peaks, and modulation spectrum for VOP Detection}\label{esm}
In this method, evidence from excitation source, spectral energy, and modulation spectrum are combined at frame level for detecting VOPs. The Hilbert envelope of LP residual contains the information about excitation source. The sum of 10 major peaks of the DFT computed for each frame, represents the energy of spectral peaks. The modulation spectrum corresponds to the gradually changing temporal envelope of speech. These methods contain different information for VOP detection and thus can be combined. The combined method leads to better performance than the excitation source, spectral energy, and modulation spectrum methods, respectively. This method is titled as COMB-ESM for the rest of the paper.

\subsection{Spectral energy around glottal closure regions for VOP Detection}\label{gci}
This method detects VOPs in a glottal closure regions of the speech signal using evidence from the spectral energy.   
The spectral energies are more prominent at GCIs. Therefore, the spectral energy is computed for the frames present in the 30\% of the glottal cycle around the GCIs. The zero frequency filter is applied for detecting the glottal closure instants in a given speech sequence. The spectral energies in the range of 500-2500 Hz are considered for VOP detection. The spectral energy signal was smoothed over the window of 50 ms to reduce the inconstancies. Further, the smoothed spectral signal is enhanced by computing the slope using first-order difference. In the enhanced signal, prominent variations (peaks) are extracted by convolving with first-order Gaussian difference operator of size 100 ms. The peaks in the convolved signal were representing the vowel onset points. This method is named as SE-GCI for the rest of the paper.

\section{Proposed VOP Detection Method}\label{prop}
 In the proposed approach, continuous wavelet transform is explored along with spectral transition measure to enhance the accuracy of the VOPs. Continuous wavelet transform can predict smooth signal features as well as abrupt transitions \cite{Reddy2017}. For some phonemes such as \textit{/ax/, /axr/ and /ux/}, CWT may fail to predict VOPs under 40 ms because of the very short and devoiced vowel. 
However, STM will accurately provide 97\% of phone boundaries within 40 ms deviation. 
For that reason, we have incorporated the information carried by STM along with CWT for further improving the performance of VOP detection. 
The details about VOP detection using CWT is included in Section \ref{cwt}. The detailed description of phone boundary detection using STM is provided in Section \ref{stm}.
The combined model for improving the performance of detected VOPs is described in Section \ref{cwt-stm}.

\begin{figure*}[t]
\centering  
\subfigure[]{\includegraphics[width=5cm, height=5cm]{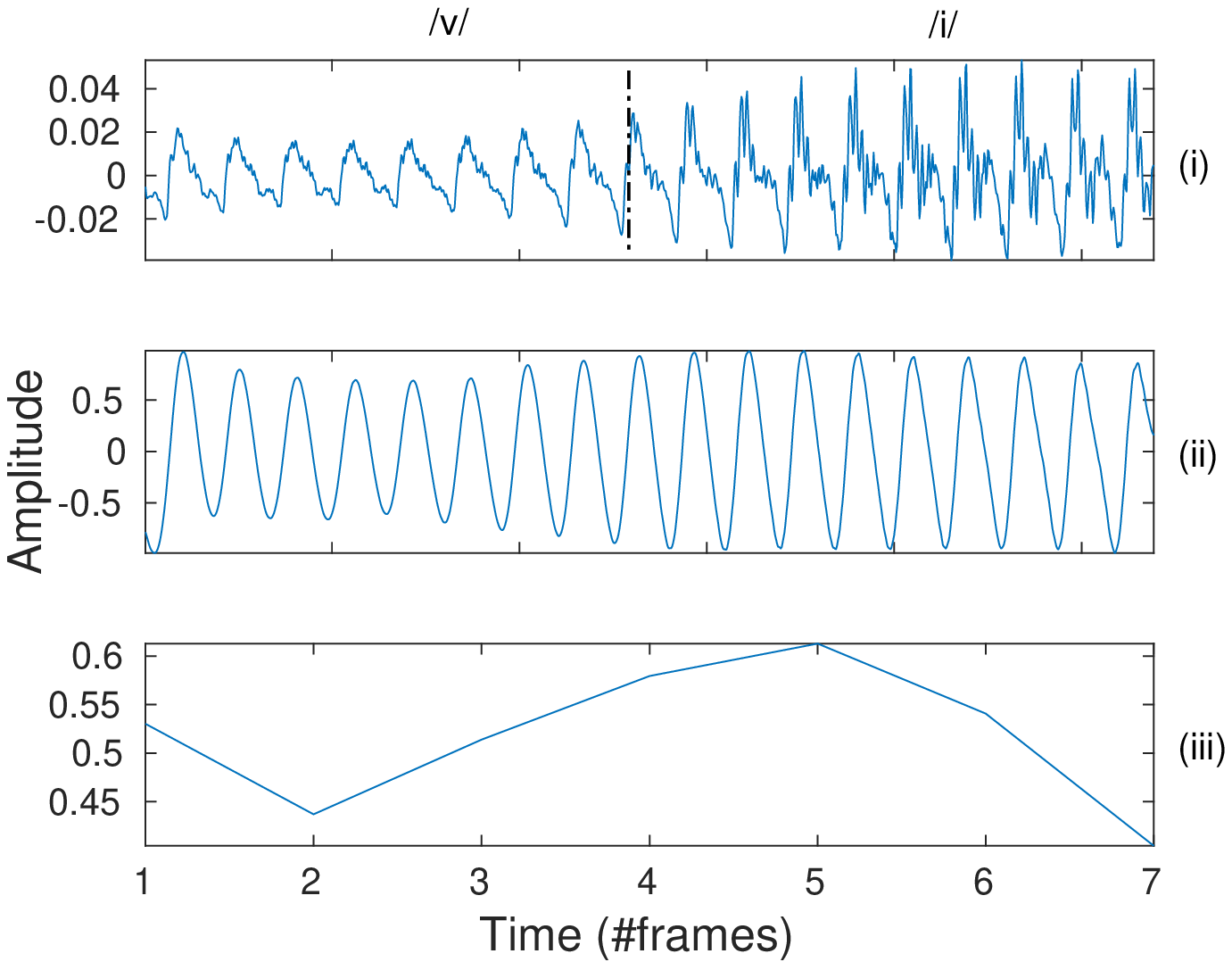}}
\subfigure[]{\includegraphics[width=5cm, height=5cm]{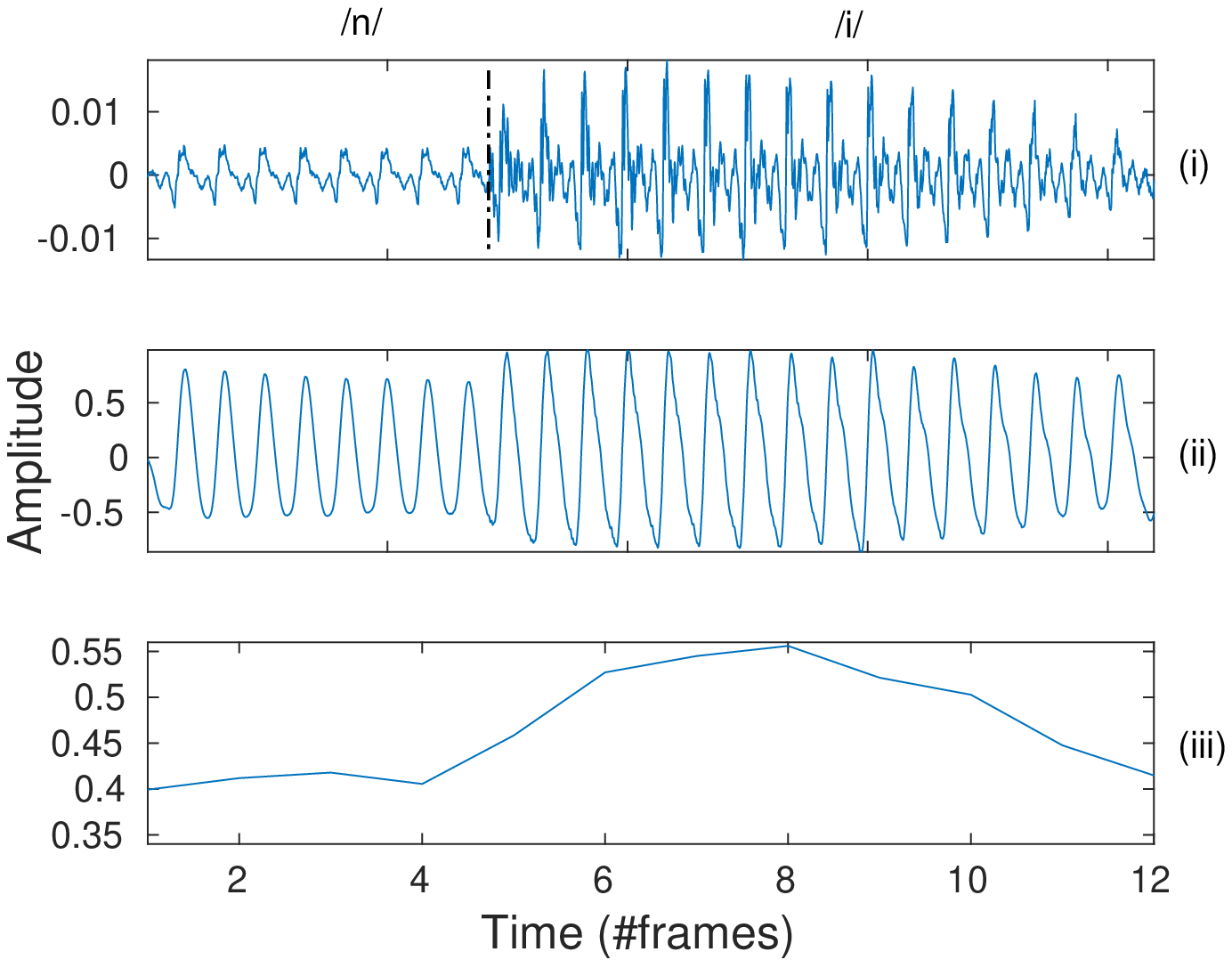}}
\subfigure[]{\includegraphics[width=5cm, height=5cm]{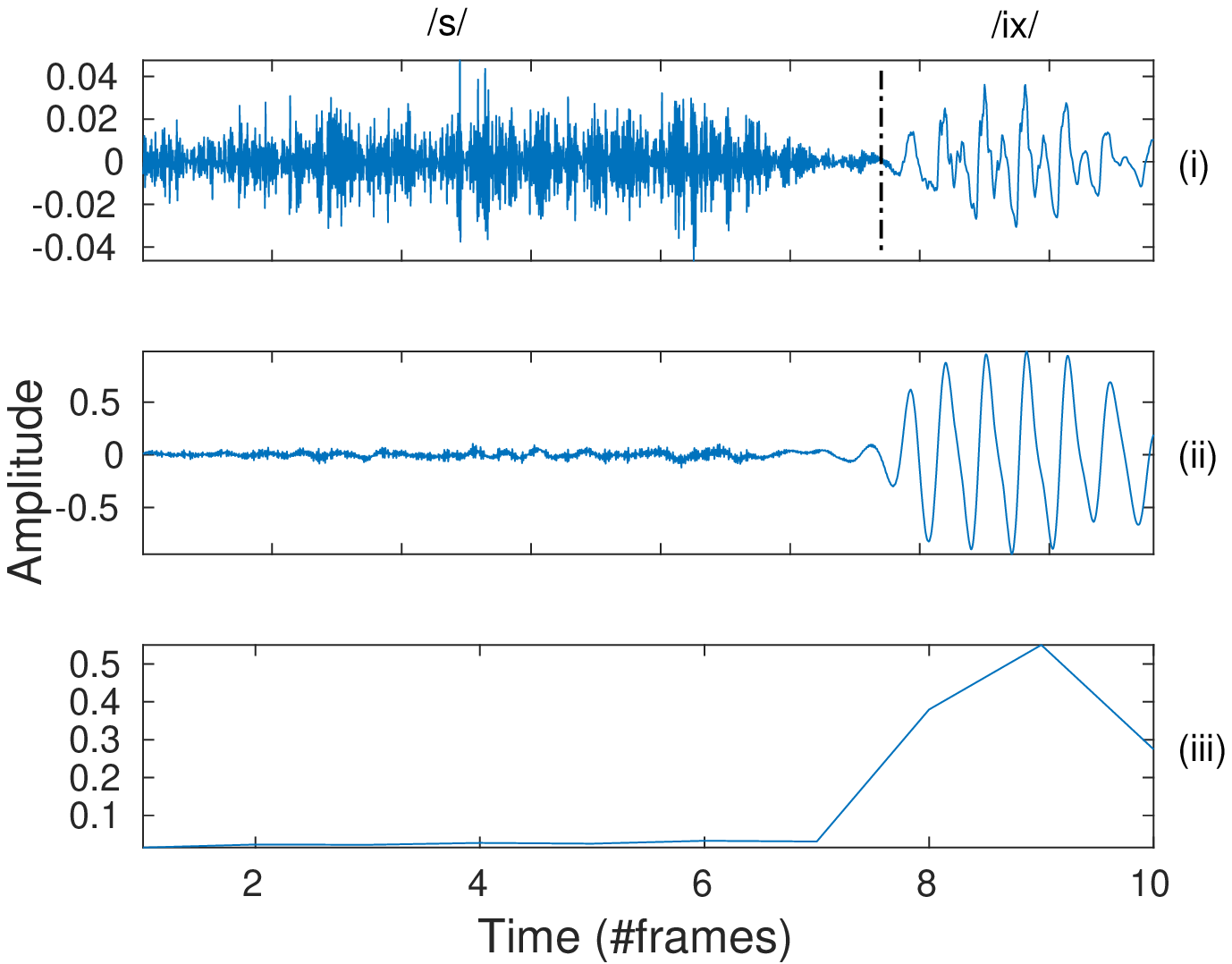}}
\subfigure[]{\includegraphics[width=5cm, height=5cm]{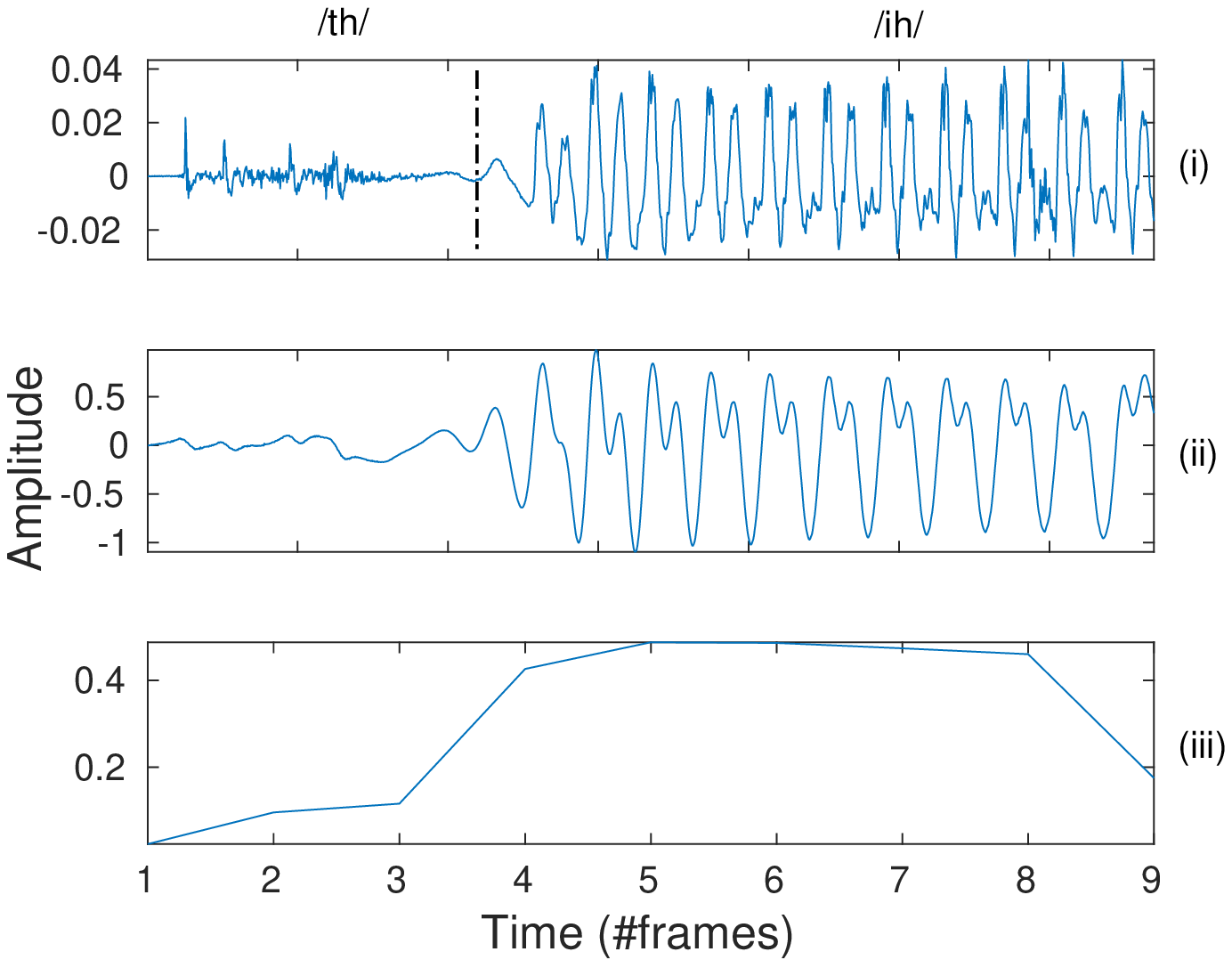}}
\caption{ Illustration of mean-signal and AAM of mean-signal. (a) Semi-vowel to vowel transition, (b) nasal to vowel transition, (c) fricative to vowel transition and (d) unvoiced to vowel transition.}\label{fig1}
\end{figure*}

\subsection{VOP Detection using CWT}\label{cwt}
CWT gives a complete representation of a signal by varying the scale value of the wavelets repeatedly. 
Mathematically, the CWT of a speech signal $x(t)$ can be represented as:
\begin{equation}
C_x(p,q)=\frac{1}{\sqrt{q}}\int_{-\infty}^{\infty}x(t)\phi^*
\Bigg(\frac{{t-p}}{q}\Bigg)dt
\end{equation}

where $\phi(t)$ is the mother wavelet and $\phi^*(t)$ is the complex conjugate of $\phi(t)$.
$C_x(p,q)$ is representing the wavelet coefficient for scale parameter $q$ $(q>0)$ and translation parameter $p$. 
The CWT coefficients computed in Eq. (1) can be observed as the  product of signal $x(t)$ and wavelet (shifted and scaled) $\phi(t): \phi_{p,q}(t)=(1/\sqrt{q})\phi({{(t-p)}}/{q})$.
In this work, VOPs are detected from the mean signal derived using CWT coefficients. The mean signal can be computed as follows:

\begin{equation}
A_c(p)=\frac{1}{N}\sum_{q \in q_s}C_x(p,q)
\end{equation}
where $N$ is the number of scales and $q_s$ is the set of chosen scale. In rest of the paper, the mean signal derived from CWT coefficients is named as ``mean-signal''. 

For detecting the VOPs, mean-signal is segmented into frames of 20 ms with 10 ms overlap. For every segment, we have calculated the average absolute magnitude (AAM) of the mean-signal. Inconsistencies in the AAM of the mean-signal are flattened by applying mean-smoothing for 40 ms segment. For determining VOPs, an optimal threshold ($th_1$) is fixed at 15\% of the maximum of smoothed AAM. This threshold is decided by estimating the errors for VOP detection. The VOP detection errors are representing the percentage of missed and spurious VOPs.
In this work, 5 distinct threshold values are examined between 11\% to 20\% at the interval of 2\% as displayed in Table \ref{error}. These experiments are performed on a subset of TIMIT speech corpus \cite{Garofalo1993}. The first column of the table contains the different threshold values, and the second column represents the percentage of missed VOPs. The third column indicates the percentage of spurious VOPs. The results shown in columns 2 and 3 are computed within 40 ms deviation.
It can be observed from Table \ref{error} that the threshold value of 15\% of the maximum of smoothed AAM is providing an appropriate decision for VOPs. In a similar manner, experiments are performed with Bengali read and conversation speech signals and analyzed that $th_1$=15\% is giving the minimum error.
As per the experimental results, threshold value of 15\% is providing the global minimum error within given data.
Therefore, in the proposed method, 15\% of the maximum of smoothed AAM is considered for determining the VOPs in TIMIT and Bengali speech signals.

\begin{table}[h]
\begin{center}
\caption{VOP detection errors for various threshold values.\label{error} }
\begin{tabular}{|c|cc|}
\hline
 {Threshold Values (\%)} &  Miss Rate & Spurious Rate\\
\hline
11 & 15 & 28  \\
13 & 15 & 25  \\
15 & 15 & 20  \\
17 & 17 & 21  \\
19 & 20 & 23  \\
\hline
\end{tabular}
\end{center}
\end{table}

 It is noticed from the literature survey that there is no work
 related to CWT for VOP detection. This motivated us to explore vowel discriminative characteristics of CWT for predicting VOPs in a speech signal. The well-known difficulties in VOP detection are finding false VOPs in case of semi-vowels, nasals, and fricatives as they are periodic in nature \cite{Prasanna2011,Kumar2016,Kumar2017}. 
 The AAM of mean-signal for semi-vowel, nasal, fricative and unvoiced speech regions are illustrated in Figure \ref{fig1}. The speech waveform and mean-signal of given speech regions are displayed in Figures \ref{fig1}(a,b,c,d)(i and ii), respectively. However, for given speech regions, the AAM of mean-signals are displayed in Figures \ref{fig1}(a,b,c,d)(iii). 
 It can be observed from the figure that the mean-signals are generally periodic in vowels, semi-vowels, and nasals, and almost zero in fricatives and unvoiced speech segments. This represents that the speech signal and mother wavelet have the least correlation at all the scales in the fricative and unvoiced regions.
 However, the CWT coefficients are higher and, shows a periodic shape at all scales for speech regions such as vowels, semi-vowels, and nasals. As we can see in Figure \ref{fig1}(a,b)(iii) that the peak amplitudes of vowel regions are significantly higher than other speech regions. Thus, the unwanted peaks can be easily removed using an optimal threshold parameter as well as using mean-smoothing. 
 Hence, it is clear that the CWT can significantly identify the VOPs in vowel regions by suppressing the remaining speech regions.  

The steps for the VOP detection using CWT based method are summarized as follows:
\begin{enumerate}
\item Compute the CWT coefficients of the speech signal.
\item Derive the mean-signal.
\item Determine AAM for each frame of the mean-signal where the length of a frame is 20 ms and frame shift is 10 ms.
\item Inconsistencies in the AAM of the mean-signal are flattened by applying mean-smoothing of frame size 40 ms.
\item Detect local peaks of the smoothed AAM of the mean-signal.
\item For removing the undesirable peaks, an optimal threshold is fixed at 15\% of the maximum of smoothed AAM. Further, undesired peaks are removed if two consecutive peaks are present within 50 ms window. On this basis, the smaller amplitude peak is removed.
\item After removing the undesired peaks, frames whose AAM is larger than or equal to the threshold value are chosen as VOP frames.
\end{enumerate}

\begin{figure}[!t]
\centering  
\includegraphics[width=3.6in]{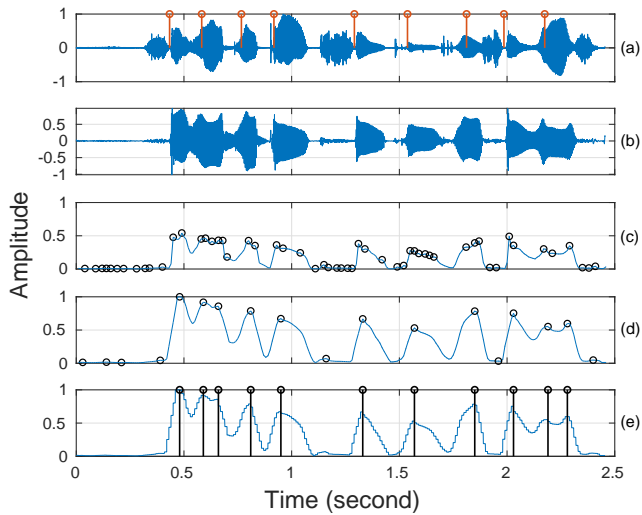}
\caption{VOP detection using CWT for an utterance \textit{/``She had your dark suit in greasy wash''/}. (a) Speech waveform with actual         VOPs, (b) Mean-signal, (c) AAM of mean-signal with peak locations, (d) Smoothed AAM with peak locations, (e) CWT detected VOPs at 15\% of the optimal threshold of the maximum of smoothed AAM. Actual VOPs are marked with the red line and detected VOPs are marked with the black line.}
\label{fig2}
\end{figure}

 Figure \ref{fig2} demonstrates the evidence of VOP detected using CWT for an utterance \textit{/``She had your dark suit in greasy wash''/} from the TIMIT corpus sampled at 16 kHz. Figure \ref{fig2}(a) displays the waveform of the speech signal. Mean-signal derived from CWT is represented in Figure \ref{fig2}(b). The average absolute magnitude (AAM) of the mean-signal is shown in Figure \ref{fig2}(c). The local peaks are represented by the circle (o) symbol. 
Figure \ref{fig2}(d), shows the smoothed AAM where mean-smoothing is applied for removing the fluctuations present in the AAM of mean-signal. The undesirable peaks in Figure \ref{fig2}(d) are omitted by applying the threshold value, which is 15\% of the maximum of smoothed AAM. 
The threshold value is empirically chosen based on several experiments on a large volume of data.
In addition to that if two consecutive peaks are reported within 50 ms; the peak with smaller magnitude will be omitted. This relies on the hypothesis that there will be only one VOP within the window of 50 ms \cite{Vuppala2012}. The detected peaks in Figure \ref{fig2}(e) after removing the undesired peaks are representing the desired locations of VOPs.

\subsection{Phone Boundary Detection using STM}\label{stm}
The spectral  transition  measure provides an unsupervised way of detecting phone boundaries in a speech signal. This is the key motivation for exploring STM in this study. 
The 13-dimensional Mel Frequency Cepstral Coefficients (MFCCs) along with $\Delta$ and $\Delta\Delta$  coefficients are considered for deriving spectral details of the speech signal. The $\Delta$ and $\Delta\Delta$ coefficients corresponds to the first and second order derivative of MFCCs, respectively. The spectral details are extracted by considering a frame size of 25 ms and a frame shift of 10 ms using the Hamming window. 

\begin{figure}[h]
\centering  \includegraphics[width=3.6in]{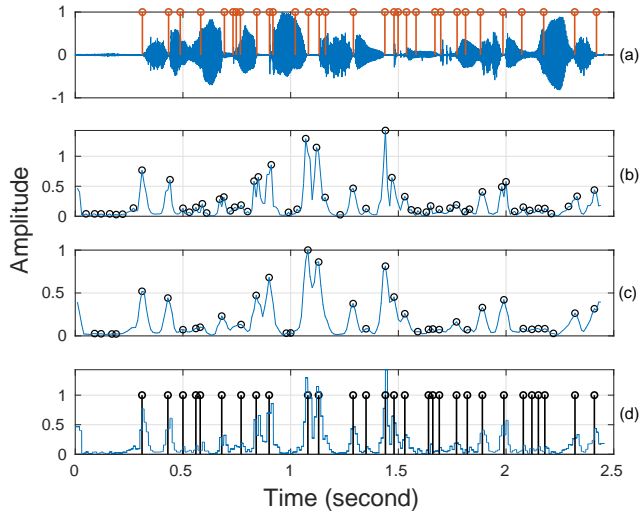}
\caption{Phone boundary detection using STM for an utterance \textit{/``She had your dark suit in greasy wash''/}. (a) Speech waveform with actual boundaries, (b) STM contour with peak locations, (c) Smoothed STM contour with peak locations,
(d) Detected phone boundaries at 12\% of the optimal threshold of the maximum of STM contour amplitude. Actual phone boundaries are marked with the red line and detected phone boundaries are marked with the black line.
\label{fig3}}
\end{figure}

The implementation of STM in this work is the same as that described in \cite{Dusan2006}. STM can be understood as the degree of variation in the spectral value of a speech signal.
The spectral variation at the phone transition is maximum compared to steady speech regions. Such higher spectral variations represent peaks and these peaks are considered as detected phone boundaries in a speech sequence. It can be observed from the STM contour (see Figure \ref{fig3}(c)) that the phone boundaries have more spectral deviation. That is responsible for producing high Mean Square Error (MSE) in linear regression \cite{Madhavi2015}.
The STM,  at  frame $g$,  can  be  computed  as  a  mean-squared value \cite{Furui1986}, i.e., 
\begin{equation}
S_g=\frac{1}{D}{\sum_{i=1}^{D}r_i^2(g)}
\end{equation}

where $S_g$ represents the STM at frame $g$, $D$  is  the  dimension  of  the  spectral  feature  vector (39  in  this  case)  and $r_i(g)$ shows the rate of variation in spectral details $MFCC_i$ and defined as \cite{Dusan2006}, 
\begin{equation}
r_i(g)=\frac{\sum_{n=-I}^{I}MFCC_i(n+g)*n}{\sum_{n=-I}^{I}n^2}
\end{equation}

where $n$ shows the frame index, $i$ is the coefficient index, and $I$ displays the number of frames (on each side of the current frame) utilized for computing the regression coefficients. The considered value of $I$ is 2 for calculating STM \cite{Furui1986}. The value of $I$ greater than 2, result in  missing desired phone locations, whereas $I$ smaller than 2, generate many unwanted phone locations.


In this work, phone boundaries are extracted from the STM contour of the speech signal. Figure \ref{fig3} illustrates the phone boundary detection for an utterance \textit{/``She had your dark suit in greasy wash''/}.
 Figure \ref{fig3}(a) represents the speech signal with actual phone boundaries. The STM contour of the speech signal is shown in  Figure \ref{fig3}(b). The local peaks are marked as a circle (o). The inconsistencies in the STM contour are flattened by applying mean-smoothing for 20 ms frame represented in Figure \ref{fig3}(c).
 The noisy peaks present in Figure \ref{fig3}(c) are removed by implementing a threshold ($th_2$) of 12\% of the maximum of STM contour amplitude. The threshold value is decided after analyzing the STM contour on a subset of TIMIT corpus. The performance of the detection task is dependant on the selection of the optimal threshold. It is noted that the threshold value smaller than $th_2$ generate more spurious boundaries, whereas the threshold value higher than $th_2$ have missed phone boundaries. 
  In Figure \ref{fig3}(b), it can be seen that the spectral value is varying significantly at phone transition. This resulted in peaks which are considered as phone boundaries in this study. The detected phone boundaries after eliminating the unwanted peaks are detected at a fixed threshold ($th_2=12$\%) is shown in Figure \ref{fig3}(d). 
The process of phone boundary detection can be summarized as follows:
\begin{enumerate}
\item Extract 39-dimensional MFCC, $\Delta$, and $\Delta\Delta$ feature, with a frame size of 25 ms and frame shift of 10 ms,
\item Compute STM for each frame of a given signal,
\item Remove inconsistencies in the STM contour by applying mean-smoothing of frame size 20 ms,
\item Detect local peaks of the smoothed STM contour,
\item For eliminating the undesirable peaks, an optimal threshold is set at 12\% of the maximum of smoothed STM contour amplitude. 
\item After removing the false peaks, frames whose amplitude is larger than or equal to the threshold value are chosen as frames for phone boundary.
\end{enumerate}

\subsection{Two-Stage Method for VOP Detection}\label{cwt-stm}
\begin{figure}[t]
\centering  \includegraphics[width=3in]{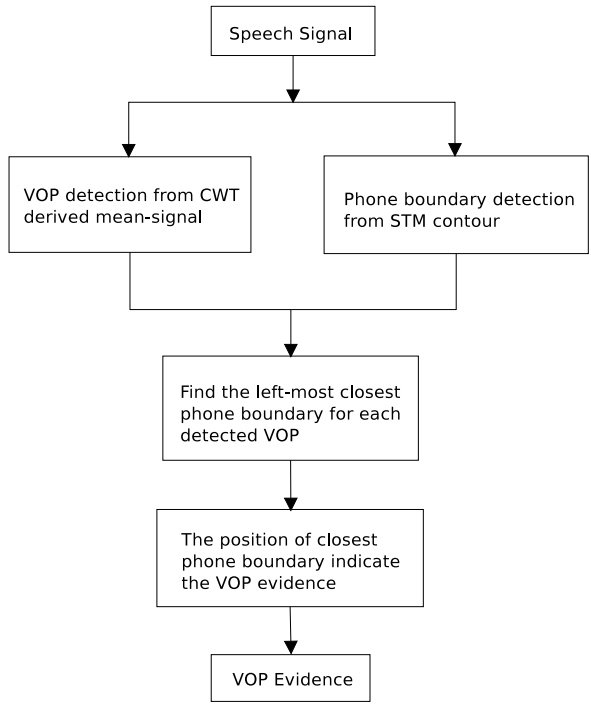}
\caption{Block diagram of proposed VOP detection method.
\label{fig5}}
\end{figure}

\begin{figure}[t]
\centering  \includegraphics[width=3.6in]{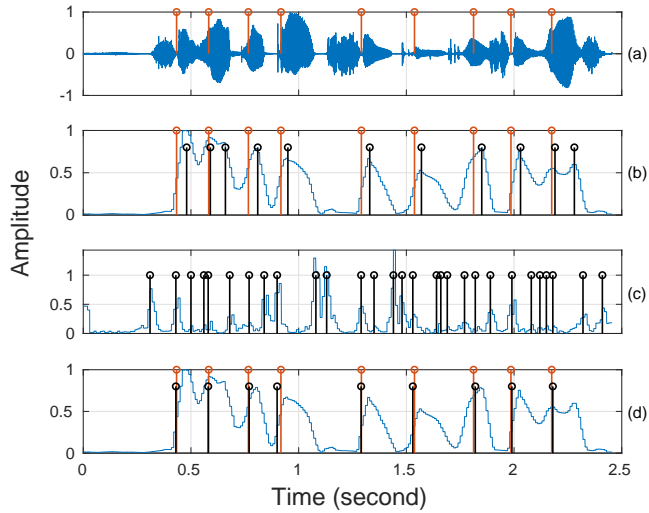}
\caption{ VOP detection for an utterance \textit{/``She had your dark suit in greasy wash''/}. (a) Speech waveform with actual VOPs, (b) Actual VOPs and detected VOPs using CWT coefficients, 
(c) Located phone boundaries using STM contour, (d) Actual VOPs and detected VOPs using the proposed method. Actual phone boundaries are marked with the red line and detected phone boundaries are marked with the black line.}
\label{fig4}
\end{figure}

 The proposed method for VOP detection is based on the evidence of two different methods discussed in Section \ref{cwt} and \ref{stm}. In the first method, VOPs are hypothesized from AAM of CWT derived mean-signal. In the second method, phone boundaries are detected from STM contour for correcting the position of CWT detected VOPs. The block diagram of the proposed VOP detection method is shown in Figure \ref{fig5}. Figure \ref{fig4} demonstrates the VOP detection for an utterance \textit{/``She had your dark suit in greasy wash''/}. The speech waveform with actual VOPs is shown in Figure \ref{fig4}(a). The detected VOPs using CWT as well as actual VOPs are depicted in Figure \ref{fig4}(b). It can be analyzed that the CWT detected VOPs are deviated from the actual VOPs. Due to this at low deviation, some actual VOPs will be accounted as missed, and some of the detected VOPs will be accounted as spurious. This will become the reason for generating missed and spurious VOPs using CWT. In addition to that, CWT will detect false VOPs in case of high energy voiced consonants; for example, in Figure \ref{fig4}(b) the detected VOPs such as 3rd and 11th are noisy.

\begin{figure}[h]
\centering  \includegraphics[width=3.6in]{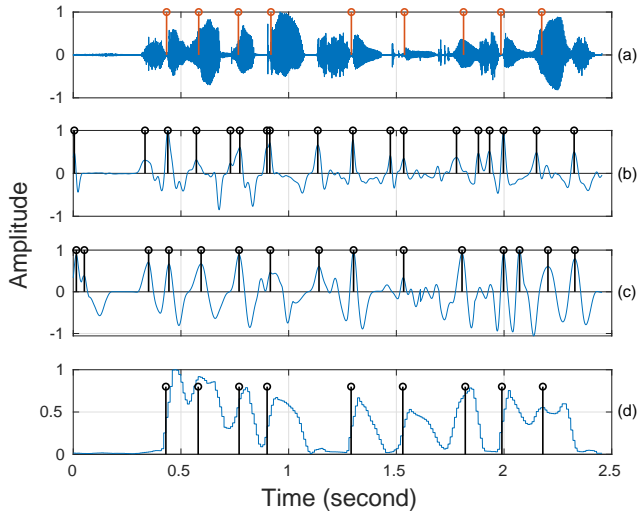}
\caption{VOP detection using proposed and existing methods for an utterance \textit{/``She had your dark suit in greasy wash''/}. (a) Speech waveform with actual VOPs, (b) Detected VOPs using COMB-ESM method, (c) Detected VOPs using SE-GCI method, (d) Detected VOPs using proposed method. Actual phone boundaries are marked with red line and detected phone boundaries are marked with black line.
\label{fig6}}
\end{figure}

\begin{table}[h]
\begin{center}
\caption{Performance of VOP detection using CWT, and STM on TIMIT corpus.\label{table1} }
\begin{tabular}{|l|cccc|c|}
\hline
{VOP} &  \multicolumn{4}{c|}{{VOPs Detected}}  &{Spurious}\\ 
{Detection} &  \multicolumn{4}{c|}{{within ms (\%)}} &{VOPs}\\ \cline{2-5}
{Method} & 10 & 20 & 30 & 40& {(\%)}\\ 
\hline
CWT & 52 & 65 & 78 & 91  & 20 \\
STM & 90 & 91 & 96 & 98  & 70 \\
\hline
\end{tabular}
\end{center}
\end{table}

Therefore, STM detected phone boundaries in Figure \ref{fig4}(c) are utilized for eliminating noisy VOPs as well as for reducing the deviation between actual and predicted VOPs. 
  Intuitively, it is reported for a subset of TIMIT speech corpus that detected VOPs are mostly occurring right side of the actual VOPs. It can also be visualized in Figure \ref{fig4}(b) that each detected VOPs have unevenly deviated towards the right side of actual VOPs.
Hence, to correct the location, the detected VOPs need to be relocated to its left side. One way to achieve this is by shifting each VOP with the fixed length. However, this is not a feasible solution because the VOPs have unevenly deviated. 
Hence, a method is required to automatically adjust the detected VOPs to their accurate locations. 
In this work, we have explored a new approach based on phone boundary details for correcting the detected VOPs. Here, the STM method is explored for phone boundary detection. 
Table \ref{table1} represents the accuracy of vowel onset point detection using CWT and STM based methods. For performing this experiment, 50 utterances are randomly selected from TIMIT speech corpus.
It is observed from Table \ref{table1} that the percentage of detected VOPs using STM is better than the CWT. However, the percentage of spurious VOPs is huge using STM than the CWT. 
 In this work, STM is evolved for phone boundary detection. Thus, at the time of VOP detection using STM, vowel boundaries are considered as VOPs, and remaining phone boundaries are considered as spurious VOPs. For that reason, the percentage of spurious VOPs are overestimated in STM based method. 
From experiments, it is noted that the performance of the proposed two-stage method does not affected by the STM detected spurious phone boundaries. 
Further, it can be seen from Table \ref{table1} that the percentage of STM detected VOPs within 10 ms and 20 ms is much higher than the CWT detected VOPs. 
This explains that STM can detect vowel boundaries (VOPs) better than CWT. Therefore, the STM based vowel boundary details are incorporated with CWT based VOP detection, for improving the performance of the proposed method at a smaller deviation.   
 The detected VOPs in Figure \ref{fig4}(d) after removing the spurious VOPs are depicting the desired location of VOPs. These VOPs are within 10 ms deviation from the actual VOPs. Hence, the proposed method is suitable for speech applications where accurate VOPs are required. 
The steps involved in correcting the detected VOPs using proposed two-stage method are: 
\begin{enumerate}
\item Detect the VOPs using the CWT based method.
\item Detect the phone boundaries using the STM based method.
\item For each detected VOP, find the left-most closest phone boundary (see in Figure \ref{fig4}(c)).
\item The position of the detected phone boundary is marked as the location of modified VOP, as shown in Figure \ref{fig4}(d).
\end{enumerate}

\section{Performance Evaluaion}\label{eval}
In this work, the performance of the proposed method is compared with two existing methods based on COMB-ESM and SE-GCI. 
Here, TIMIT corpus is considered for evaluating the performance of VOP detection methods. However, Indian speech corpora in read and conversation modes are considered for depicting the significance of proposed VOP detection method.

\subsection{VOP Detection on TIMIT Corpus}
 Experiments are conducted on TIMIT speech corpus for performance analysis of proposed and two existing VOP detection methods. About 120 randomly selected utterances sampled at 16 kHz are used for analyzing the performance of the explored VOP detection methods. 
The metrics considered for measuring the performance of the various methods are identification rate (IR), average deviation (AD), missing rate (MR), and spurious rate (SR). The percentage of actual VOPs that correspond to the detected VOPs within the considered (10-40 ms) time-resolutions is known as identification rate. The average deviation (in ms) is demonstrating the average time difference between the actual and predicted VOPs. The percentage of actual VOPs that are undetected within the considered deviation is termed as missed rate. The percentage of detected VOPs other than the actual VOPs is termed as spurious rate. 

VOP detection accuracy of proposed, COMB-ESM and SE-GCI methods in terms of IR, AD, MR, and SR is demonstrated in Table \ref{table2}. The first column contains the list of methods involved in analyzing the VOP detection performance. Columns second to fifth represent the IR (\%) within the mentioned deviations. The sixth column specifies the AD (in ms) with respect to the actual VOPs. Seventh and eighth columns, represent the missed and spurious rates, respectively.
It is observed from the table that the overall performance of the proposed method is better than the existing (COMB-ESM, and SE-GCI) methods for VOP detection. The average deviation in the proposed method (7 ms) is significantly smaller than the COMB-ESM (18 ms) and SE-GCI (13 ms) methods. The rate of missed and spurious VOPs is relatively higher in COMB-ESM and SE-GCI methods (see Figure \ref{fig6}). 
In both COMB-ESM and SE-GCI, the detection of VOPs relies upon the spectral energy and its enhancement. In these methods, the spectral energy of a speech signal is enhanced by computing  its slope value using first-order derivative (FOD). Further, the enhanced features are convolved with first order Gaussian difference (FOGD) operator for locating the VOP evidences. 
Here, enhanced feature help in improving the identification rate  but at the same time it highlighted the peaks for periodic non-vowel regions such as semi-vowels, and nasals, which leads to spurious detection of VOPs. 
The proposed method outperformed the existing methods in case of identification and spurious rates. It can be seen in Figure \ref{fig6} that the existing spurious VOPs in COMB-ESM and SE-GCI methods are removed in the proposed method.
Additionally, it is noticed that the IR for the proposed method is almost 30\% higher within 10 ms as compared to other existing methods. However, the proposed method is shown the significantly better performance within 20 ms than the 10 ms deviation.
This is due to the cases where high energy voiced consonants are preceded by the vowels, which resulted in the deviation of detected VOPs with respect to the genuine VOPs.

\begin{table}[t]
\begin{center}
\caption{Performance of VOP detection using proposed method, COMB-ESM and SE-GCI on TIMIT corpus.\label{table2} }
\begin{tabular}{|l|cccc|c|c|c|}
\hline
{VOP} &  \multicolumn{4}{c|}{{VOPs Detected}} & {Average} & {Missed} &{Spurious}\\
{Detection} &  \multicolumn{4}{c|}{{within ms (\%)}}& {Deviation} & {VOPs} &{VOPs}\\ \cline{2-5}
{Method} & 10 & 20 & 30 & 40& {($\approx$ ms)} & {(\%)} &{(\%)}\\ 
\hline
COMB-ESM  & 51 & 59 & 74 & 90& 18 & 10&6 \\
SE-GCI  & 62 & 79 & 86 & 91& 13 & 9 & 5 \\
Proposed & 82 & 88 & 91 & 92& 7 & 8 & 3 \\
\hline
\end{tabular}
\end{center}
\end{table}

\subsection{VOP Detection in Read and Conversation Modes of Bengali Speech Corpora}


 In this work, the significance of the proposed method is demonstrated by detecting the VOPs in read and conversation modes of Bengali speech.  
The Bengali speech dataset is collected as part of consortium project titled \textit{Prosodically guided phonetic engine for searching speech databases in Indian languages} supported by DIT, Govt. of India \cite{Kumar2013}. In this study, read speech is collected from news reading, and the conversational speech is collected from casual talks. 
The speech signals are sampled at a rate of 16 kHz with the precision of 16 bits per sample.
Altogether, 20 utterances are collected from 5 distinct speakers, where 3 male and 2 female speakers are considered for each mode. 
About 100 utterances from each mode are selected for evaluating the performance of the proposed and existing methods. 

Mostly, VOP detection methods are studied for read mode of speech \cite{Hermes1990,Gangashetty2004,Gangashetty2004a,Rao2009,Prasanna2009, Vuppala2012}. However, speech can be broadly divided into two modes, such as read and conversation. The acoustic and linguistic characteristics of these modes are very different. As conversation speech is a type of spontaneous and unconstrained communication between two or more than two people. However, the read speech includes planning before reading in constrained conditions such as news reading. The conversation mode includes higher variations in activity of vocal folds than the read mode. Due to the aforementioned variations, read, and conversation modes are examined in this work for evaluating the accuracy of VOP detection methods.

\begin{figure}[t]
\centering { \includegraphics[width=3.7in]{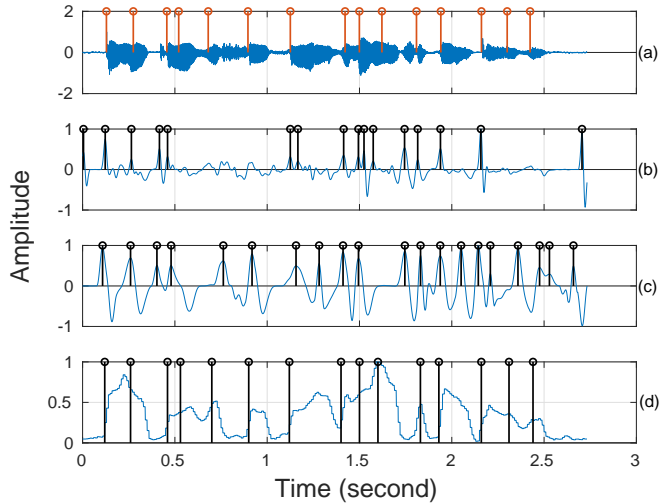}}
\caption{ VOP detection for Bengali sentence \textit{/``Tomake abosi fael gulo muche felte hobe''/} uttered in read mode. (a) Speech waveform with actual VOPs, (b) Located VOPs using COMB-ESM method, (c) Located VOPs using SE-GCI method, (d) Detected VOPs using the proposed method. Actual phone boundaries are marked with the red line and detected phone boundaries are marked with the black line.}
\label{fig7}
\end{figure}

\begin{figure}[t]
\centering { \includegraphics[width=3.6in]{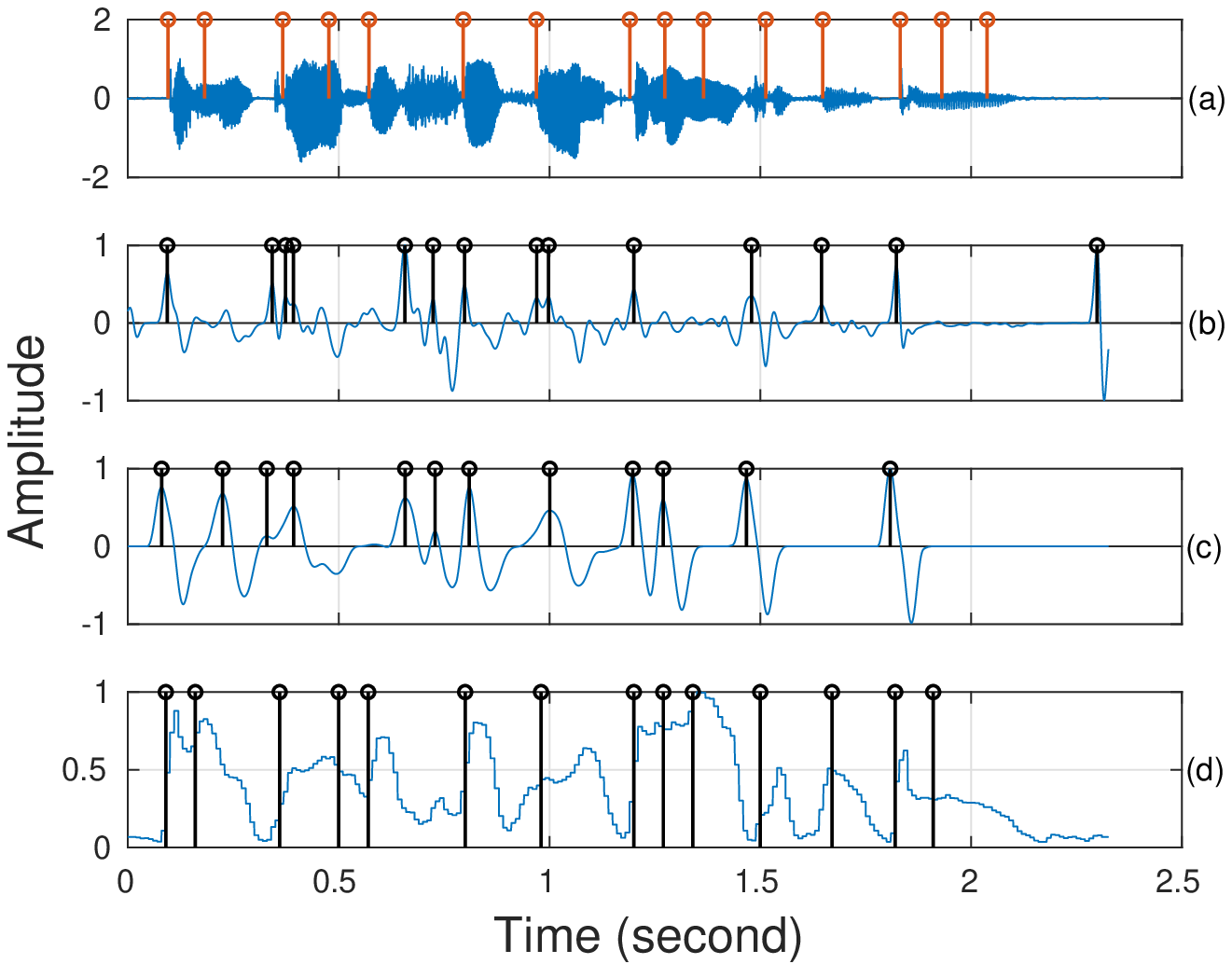}}
\caption{ VOP detection for Bengali sentence \textit{/``Tomake abosi fael gulo muche felte hobe''/} uttered in conversation mode. (a) Speech waveform with actual VOPs, (b) Located VOPs using COMB-ESM method, (c) Located VOPs using SE-GCI method, (d) Detected VOPs using the proposed method. Actual phone boundaries are marked with the red line and detected phone boundaries are marked with the black line.}
\label{fig8}
\end{figure}

\begin{table}[!t]
\begin{center}
\caption{Performance of VOP detection using COMB-ESM, SE-GCI, and proposed methods on read and conversation (Conv) modes of Bengali speech corpora.\label{table3} }
\begin{tabular}{|l|l|cccc|c|c|}
\hline
Speech&{VOP} &  \multicolumn{4}{c|}{{VOPs Detected}} & {Average} &{Spurious}\\
Mode&{Detection} &  \multicolumn{4}{c|}{{within ms (\%)}}& {Deviation} &{VOPs}\\ \cline{3-6}
&{Method} & 10 & 20 & 30 & 40& {($\approx$ ms)} &{(\%)}\\ 
\hline
\multirow{3}{*}{Read}
&COMB-ESM & 44 &63 &78&87&20&8\\
&SE-GCI  & 59 &70 &82 &89 &14 &5 \\
&Proposed & 75 &81 &88 &89&13 &4 \\
\hline	
\multirow{3}{*}{Conv}
&COMB-ESM & 40 &46 &61 &73&21 &15\\
&SE-GCI   & 45 &58 &72 &78&19 &13 \\
&Proposed & 72 &79 &83 &85&15 &7 \\
\hline
\end{tabular}
\end{center}
\end{table}

Table \ref{table3} demonstrates the performance of VOPs detected in read and conversation modes of Bengali speech using the existing and proposed methods. The first column shows the modes of speech considered for detecting VOPs. The second column signifies various methods used in the analysis of VOP detection. 
Third, to sixth columns represent the IR (\%) within the given deviations (10 to 40 ms). Seventh and eighth columns are specifying the average deviation and spurious rates, respectively. 

It is noticed from Table \ref{table3} that the performance of detected VOPs is higher under read mode compared to conversation mode. In conversation mode, percentage of detected VOPs is reduced by 14\%, 11\% and 4\% (within 40 ms deviation) as compared to read mode using COMB-ESM, SE-GCI and proposed method, respectively. It is important to note that the performance reduction in conversation mode is minimum for the proposed method than the existing methods. Hence, this result states that the proposed method is least influenced by the speech mode discriminative characteristics.
This analysis can be visualized in Figures \ref{fig7} and \ref{fig8} where, a Bengali sentence \textit{/``Tomake abosi fael gulo muche felte hobe''/} is spoken in read and conversation modes, respectively. 
For both the modes, the same female speaker has uttered the given sentence. The speech recorded in conversation mode includes fear expression. However, uttered speech in read mode contains a neutral expression. 
Here, Figures \ref{fig7}(a) and \ref{fig8}(a) show the speech waveform with actual VOPs in read and conversation modes, respectively. The detected VOPs using COMB-ESM method in read and conversation modes are respectively depicted in Figures \ref{fig7}(b) and \ref{fig8}(b). Figures \ref{fig7}(c) and \ref{fig8}(c) show the detected VOPs using SE-GCI method in read and conversation modes, respectively. Figures \ref{fig7}(d) and \ref{fig8}(d) represent the identified VOPs using the proposed method in read and conversation modes, respectively.
It is observed from the Figures \ref{fig7} and \ref{fig8}, the energy variation is highly dynamic in conversation mode than the read mode. 
 This observation is justified intuitively by analyzing the Bengali sentences in given modes. Further, it is claimed that the spectral energy is significantly varying in conversation mode than the read mode due to the presence of emotions.
The involvement of emotions leads to a small percentage of clean speech in conversation mode than the read mode of speech. Therefore, the percentage of detected VOPs are improved in read mode than the conversation mode.
In addition to that, while expressing emotions, some voiced consonants are also got emphasized in conversation mode, which results in spurious VOPs. It is observed that the duration of vowels is smaller in conversation mode as compared to read mode. This is the reason for missing VOPs in conversation mode. The paralinguistic aspects of the speech, such as gasp, sigh, and mhm are more often present in conversation speech than the read speech. This resulted in spurious detection of VOPs in conversation mode.
For all these reasons, the overall performance of VOP detection methods is reduced in conversation mode as compared to read mode. 

It can be noted from Table \ref{table3} that proposed method is performing better than the existing methods for spotting VOPs in case of read and conversation modes.
In read mode, the proposed method is detecting about 17\% more VOPs under 10-20 ms deviation as compared to COMB-ESM and SE-GCI. Similarly, in conversation mode, the proposed method is extracting almost 35\% more VOPs within 10-20 ms deviation than the COMB-ESM and SE-GCI. 
The complete result represents that the performance of the proposed method is relatively similar for read and conversation speech. 
The instants of VOPs represent sharp energy transitions in both read and conversation speech. The ability of CWT to confine these sharp energy transitions, help in accurately detecting VOPs even in the presence of conversation speech. Hence, the proposed method is providing better identification rate for both the conversation and read modes of speech.
Further, it can be noticed that the proposed method has a significant reduction in the average deviation and spurious rates.
Percentage of spurious VOPs is reduced around 4\% and 8\% in the proposed method as compared to COMB-ESM for read and conversation modes, respectively. Similarly, the percentage of spurious VOPs using the proposed method as compared to SE-GCI is reduced around 1\% and 6\% in read and conversation modes, respectively. 
This is because, at first step in the proposed method, AAM of mean-signal obtained from CWT coefficients is enhancing the voiced region and suppressing the unvoiced region of speech.
At second step, the detected VOPs are corrected using phone boundaries derived from STM contour. The use of STM in proposed method helps in removing the spurious VOPs as well as in reducing the deviation between actual and predicted VOPs. 
The overall result indicates that combining CWT and STM methods into a single framework can accurately detect the VOPs present in a speech utterance spoken in any mode.

\section{Conclusion}\label{conc}
In this work, a novel method is proposed for accurate VOP detection.
The proposed method consists of two-stages. At the first stage, VOPs are detected by using the AAM of mean-signal derived from continuous wavelet transform (CWT) coefficients. At the second stage, the evidence of identified VOPs is corrected with the presence of the nearest phone boundary detected using spectral transition measure (STM) method. In the proposed method, CWT and STM are utilized to obtain the sharp energy transitions around the VOPs. 
VOP detection experiments are carried out with TIMIT corpus (read speech) and Bengali corpus (read and conversation speech). Performance of the proposed approach is compared with two standard methods: COMB-ESM and SE-GCI.
The proposed method was demonstrated to be significantly better in eliminating spurious VOPs and for accurately detecting the VOPs within 10 ms deviation as compared to COMB-ESM and SE-GCI methods. The efficiency of the proposed method is demonstrated by detecting VOPs in two acoustically and linguistically different speech modes such as read and conversation modes. 
The results achieved for read and conversation modes signify that the performance of the proposed method is insignificantly affected by the acoustic variation among the modes and achieved almost similar performance for both the modes.
As the proposed approach demonstrates accurate computation of VOP locations in a speech signal, this can be utilized for consonant-vowel recognition, speech rate modification, voiced-unvoiced classification, and so on.
Further, the robustness of the proposed method can be explored for noisy speech corpora. In this work, we have explored two broad modes of speech for examining the significance of the proposed method; one can explore other emotional modes of speech like anger, happy, sad, etc. The proposed method is explored for VOP detection and in future, it may be examined for detecting the vowel end points (VEPs) in read and conversation modes.


\bibliographystyle{elsarticle-num}
\bibliography{vop_bib}


%
%



\end{document}